\documentclass[12pt]{article}

\usepackage[english]{babel}

\usepackage[english]{babel}
\usepackage[dvips, final]{graphicx}
\usepackage[T1]{fontenc} 
\usepackage{amsmath}
\usepackage{amsfonts}
\usepackage{amssymb}
\usepackage{amsthm}
\usepackage{enumerate}
\usepackage{color}
\usepackage{epsfig}

\usepackage{makeidx}
\usepackage{xspace}
\usepackage{tabularx}
\usepackage{verbatim}
\usepackage[14pt]{extsizes}
\usepackage{geometry}
\usepackage{texnames}
\usepackage{fancyhdr}
\usepackage{cite}
\graphicspath{{:fig:}}

\RequirePackage{amsthm}
\theoremstyle{plain} 

\usepackage{amsmath, amsthm, amscd, amsfonts, amssymb, graphicx, color}

\newtheorem{theorem}{Theorem}[section]
\newtheorem{lemma}[theorem]{Lemma}

\newtheorem{corollary}[theorem]{Corollary}
\theoremstyle{definition}
\newtheorem{definition}[theorem]{Definition}

\theoremstyle{remark}
\newtheorem{remark}[theorem]{Remark}
\numberwithin{equation}{section}
\usepackage{tabularx}

\usepackage{cite}

\usepackage{dcolumn}
\usepackage{bm}

\theoremstyle{remark}

\def\@tocrmarg{3em}
\def\@pnumwidth{2em}

\begin{document}
\begin{center}
 {\large \bf On Application of the Lyapunov-Schmidt-Trenogin Method to Bifurcation Analysis 
of the Vlasov-Maxwell system}
\end{center}

\begin{center}
Nikolai  A. Sidorov
\end{center}
\begin{center}
$^{1}$Irkutsk State University, Karl-Marks Str. 1, \\
664003 Irkutsk, Russia\\

\end{center}

\begin{abstract}
The review of existence theorems of bifurcation points of solutions for nonlinear
operator equation in Banach spaces is presented.
 The sufficient conditions of bifurcation of solutions of boundary-value
problem for Vlasov-Maxwell system are considered. The analytical method of 
Lyapunov-Schmidt-Trenogin is employed.
\end{abstract}


%
%
%
%
%
%
%

$\,$\\

\noindent

\section*{Introduction}

One of the substantial  problems in  Plasma Physics  is  study of  kinetic
Vlasov-Maxwell (VM) system \cite{20} describing a behaviour of many-component collisionless plasma.
The bibliography on the existence of solutions for the VM system is
available, for example, see under references \cite{1, 3, 11, 21, 35} and the
references given there. Nevertheless, the problem of
bifurcation analysis of VM system, which was first formulated by
A.A.Vlasov \cite{20}, has appeared very complicated on the background of progress of
bifurcation theory in other fields and it remains open up to the present time.
There are only some isolated results.
In \cite{9, 10} the VM system is reduced to the system of semilinear elliptical
equations for special classes of distribution functions introduced in
 \cite{12}. The relativistic version of VM system for such distributions was considered
in \cite{1}. One simple existence theorem of a point of bifurcation is announced in \cite{13},
and another one is proved for this system in \cite{14}.

Vladilen A. Trenogin laid out the fundamentals of the modern analytical
branching theory of nonlinear equations. Here readers may refer to his monograph
\cite{19}, chapters 7-10. The bifurcation theory have been developed 
by various authors \cite{5,6,15,16,17,18, 21, 22, 25, 30}, etc.
The approximate methods of construction branching solutions were
constructed in \cite{21, 22, 24, 25, 26, 27, 28, 30, 31, 32, 33, 34}, \cite{36,37,38,39}.
The readers may refer to the pioneering research contributions presented in original paper \cite{13, 14, 23, 29}, in the
monograph \cite{21} as well as in recently published monograph \cite{35} in
the field of bifurcation analysis of the Vlasov-Maxwell systems.

The objective of the present paper is to give the survey of  a general existence theorems of
bifurcation points of VM system with the given boundary conditions on potentials
of an electromagnetic field both the densities of charge and current. Here we
apply our results of bifurcation theory from \cite{15, 17, 21, 22, 23} and we use the index
theory \cite{2, 7, 22} for the study of bifurcation points of the VM system.

We consider the many-component plasma consisting of electrons and positively
charged ions of various species, which described by the many-particle distribution function
$f_{i}=f_{i}(r,v)$, $i=\overline{1,N}$. The plasma is confined to a domain
$D\subset {\mathbb{R}}^{3}$ with smooth boundary. The particles are to interact only by
self-consistent force fields, collisions among particles being neglected.

The behaviour of plasma is governed by the following version of the stationary
VM system \cite{20}
$$
v\cdot\partial_{r}f_{i}+q_{i}/m_{i}(E+\frac{1}{c}v\times B)\cdot\partial_{v}
f_{i}=0,  \eqno(1)
$$
$$
r\in D\subset R^{3}, \;\;i=1,\ldots,N, \hspace{1.2cm}
$$
$$
curl E=0,
$$
$$
div  B=0
$$
$$
divE=4\pi\sum_{k=1}^{N}q_{k}\int_{R^{3}}f_{k}(r,v){\rm d}v\stackrel{\triangle}{=}\rho,
\hspace{0.6cm}       \eqno(2)
$$
$$
curlB=\frac{4\pi}{c}\sum_{k=1}^{N}q_{k}\int_{R^{3}}vf_{k}(r,v){\rm d}v\stackrel
{\triangle}{=}j.
$$
Here $\rho(r),\;j(r)$ are the densities of charge and current, and $E(r),\;B(r)$
are the electrical and the magnetic fields.

We seek the solution $E,\;B,\;f$ of VM system (1)-(2) for $r\in D\subset {\mathbb{R}}^{3}$
with boundary conditions on the potentials and the densities
$$
U\mid_{\partial D}=u_{01}, \;\;\;(A,d)\mid_{\partial D}=u_{02}; \eqno(3)
$$
$$
\rho\mid_{\partial D=0},\;\;\;\;j\mid_{\partial D}=0, \eqno(4)
$$
where $E=-\partial_{r}U$, $B=curl A$, and $U,\;A$ be scalar and vector potentials.

We call a solution $E^{0},\;B^{0},\;f^{0}$ for which $\rho^{0}=0$ and
$j^{0}=0$ in domain $D$, trivial.

In the present paper we investigate the case of distribution functions of the
special form \cite{9}
$$
f_{i}(r,v)=\lambda\hat{f}_{i}(-\alpha_{i}v^{2}+\varphi_{i}(r),\;v\cdot d_{i}+
\psi_{i}(r))\stackrel{\triangle}{=}\lambda\hat{f}_{i}({\bf R},{\bf G})
\eqno(5)
$$
$$
\varphi_{i}: R^{3}\rightarrow R;\;\;\;\psi_{i}: R^{3}\rightarrow R;\;\;\;
r\in D\subseteq R^{3};\;\;\;v\in R^{3};
$$
$$
\lambda\in R^{+};\;\;\;\alpha_{i}\in R^{+}\stackrel{\triangle}{=}[0,\infty);\;\;\;
d_{i}\in R^{3}, \;\;i=1,\ldots,N,
$$
where functions $\varphi_{i},\;\psi_{i}$, generating the appropriate electromagnetic
field $(E,\;B)$, has to be defined.

We are interested in the dependence of unknown functions $\varphi_{i},\;\psi_{i}$
upon parameter $\lambda$ in distribution (5). Here we study the case, when $\lambda$
in (5) does not depend on physical parameters $\alpha_{i}$  and $d_{i}$.
The general case of a bifurcation problem with
$\alpha_{i}=\alpha_{i}(\lambda)$, $d_{i}=d_{i}(\lambda)$,
$\varphi_{i}=\varphi_{i}(\lambda,r)$, $\psi_{i}=
\psi_{i}(\lambda,r)$ will be considered in the following paper.

\begin{definition} 
The point $\lambda^{0}$ is called a  bifurcation point of the
solution of VM system with conditions (3), (4), if in any neighbourhood of
vector $(\lambda^{0},E^{0},B^{0},f^{0})$, corresponding to the trivial solution with
$\rho^{0}=0$, $j^{0}=0$ in domain $D$, there is a vector $(\lambda,E,B,f)$
satisfying to the system (1)-(2) with (3), (4) and for which
$$
\parallel E-E^{0}\parallel+\parallel B-B^{0}\parallel+\parallel f-f^{0}
\parallel>0.
$$
\end{definition}

Let $\varphi_{i}^{0},\;\psi_{i}^{0}$ are such constants that the corresponding
$\rho^{0}$ and $j^{0}$, induced by distributions $f_{i}$ in the medium for
$\varphi_{i}^{0},\;\psi_{i}^{0}$, are equal to zero in domain $D$. Then
VM system has the trivial solution
$$
f_{i}^{0}=\lambda\hat{f}_{i}(-\alpha_{i}v^{2}+\varphi_{i}^{0},\;v\cdot d_{i}+
\psi_{i}^{0}), \;\;\;E^{0}=0,\;\;\;B^{0}=\beta d_{1},\;\;\;\beta-const\;\;
{\rm for}\;\;\forall\lambda.
$$

The organization of the present work is as follows. In Section  2 two theorems of existence of bifurcation points for the nonlinear
operator equation in Banach space generalizing known results on a bifurcation point
 are proved. The method of proof of these theorems uses the index theory
of vector fields \cite{2, 7} and allows to investigate not only the point, but also
the bifurcation surfaces with minimum restrictions on equation.

In Section 3 we reduce the problem on a bifurcation point of VM system to
the problem on bifurcation point of semilinear elliptic system.
Last one is treated as the operator equation in Banach space. We derive the branching
equation (BEq) which allows to prove the principal theorem of existence of bifurcation
points of VM system because of results of the Section 2. An essential moment here is
that the semilinear system of elliptic equations is potential that reduces to
potentiality of BEq.

It follows from our results that for the original problem (1) -- (4) the bifurcation is possible
only in the case, when number of species of particles $N\ge 3$.

\section{Bifurcation of solutions of nonlinear equations in Banach spaces}

Let $E_{1},\;E_{2}$ are real Banach spaces; $\Upsilon$
be normalized space. Consider the equation
$$
Bx=R(x,\varepsilon). \eqno(6)
$$
Here $B: D\subset E_{1}\rightarrow E_{2}$ be closed linear operator with a dense range
of definition in $E_{1}$. The operator $R(x,\varepsilon)$ with values in
$E_{2}$ is defined, is continuous and continuously differentiable by Frechet
with respect to $x$ in a neighbourhood
$$
\Omega=\{x\in E_{1},\;\varepsilon\in\Upsilon:\parallel x\parallel<r,\;\parallel
\varepsilon\parallel<\varrho\}.
$$
Thus, $R(0,\varepsilon)=0$, $R_{x}(0,0)=0$. Let operator $B$ be Fredholm.
Let us introduce the basis $\{\varphi_{i}\}_{1}^{n}$ in a subspace $N(B)$, the basis $\{
\psi_{i}\}_{1}^{n}$ in $N(B^{*})$, and also the systems $\{\gamma_{i}\}_{1}^{n}\in
E_{1}^{*}$, $\{z_{i}\}_{1}^{n}\in E_{2}$ which are biorthogonal to these basises.

\begin{definition}
 The point $\varepsilon_{0}$ is called a bifurcation point of the
equation (6), if in any neighbourhood of  point $x=0,\;\varepsilon_{0}$
there is a pair $(x,\varepsilon)$ with $x\neq 0$ satisfying to the equation (6).
\end{definition}

It is well known \cite{19} that the problem on a bifurcation point of
(6) is equivalent to the problem on bifurcation point of finite-dimensional
system
$$
L(\xi,\varepsilon)=0, \eqno(7)
$$
where $\xi\in R^{n}$, $L: R^{n}\times\Upsilon\rightarrow R^{n}$. We call equation
(7) the branching equation (BEq). We wright (6) as the system
$$
\tilde{B}x=R(x,\varepsilon)+\sum_{s=1}^{n}\xi_{s}z_{s}  \eqno(8)
$$
$$
\xi_{s}=<x,\gamma_{s}>,\;\;s=1,\ldots,n, \eqno(9)
$$
where $\tilde{B}\stackrel{def}{=}B+\sum_{s=1}^{n}<\cdot,\gamma_{s}>z_{s}$
has inverse bounded. The equation (8) has the unique small solution
$$
x=\sum_{s=1}^{n}\xi_{s}\varphi_{s}+U(\xi,\varepsilon) \eqno(10)
$$
at $\xi\rightarrow 0$, $\varepsilon\rightarrow 0$. Substitution (10) into (9)
yields formulas for the coordinates of  vector-function $L:R^{n}\times\Upsilon\rightarrow
R^{n}$
$$
L_{k}(\xi,\varepsilon)=<R\biggl (\sum_{s=1}^{n}\xi_{s}\varphi_{s}+U(\xi,
\varepsilon),\varepsilon\biggl ),\psi_{k}>. \eqno(11)
$$
Here derivatives
$$
\frac{\partial L_{k}}{\partial\xi_{i}}\mid_{\xi=0}=<R_{x}(0,\varepsilon)
(I-\Gamma R_{x}(0,\varepsilon))^{-1}\varphi_{i},\psi_{k}>\stackrel{def}{=}
a_{ik}(\varepsilon)
$$
are continuous in a neighbourhood of point $\varepsilon=0$, $\parallel\Gamma R_{x}(0,
\varepsilon)\parallel<1$.

Let us introduce a set $\Omega=\{\varepsilon\mid\det[a_{ik}(\varepsilon)]=0\}$,
containing  point  $\varepsilon=0$ and the following condition:

{\bf A)} Suppose that in a neighbourhood of  point $\varepsilon_{0}\in\Omega$ there is
a set $S$, being Jordan continuum, representable as  $S=S_{+}\bigcup S_{-}$,
$\varepsilon_{0}\in \partial S_{+}\bigcap \partial
S_{-}$. Moreover, there is a continuous map $\varepsilon(t)$,
$t\in [-1,1]$ such that $\varepsilon:[-1,0)\rightarrow S_{-}$, $\varepsilon:
(0,1]\rightarrow S_{+}$, $\varepsilon(0)=\varepsilon_{0}$, $\det[a_{ik}(
\varepsilon(t))]_{i,k=1}^{n}=\alpha(t)$, where $\alpha(t): [-1,1]\rightarrow R^{1}$
be continuous function vanishes only at $t=0$.

\begin{theorem} { Assume condition {\bf A}, and $\alpha(t)$
is monotone increasing function. Then $\varepsilon_{0}$ be a bifurcation point of
(6)}.
\end{theorem}

Proof. We take arbitrarily small $r>0$ and $\delta>0$.
Consider the continuous vector field
$$
H(\xi,\Theta)\stackrel{def}{=}L(\xi,\varepsilon((2\Theta-1)\delta)): R^{n}\times
R^{1}\rightarrow R^{n},
$$
defined at $\xi, \Theta\in M$, where $M\{\xi,\Theta\mid\parallel\xi
\parallel=r,\;0\leq\Theta\leq 1\}$.

Case 1. If there is a pair $(\xi^{*},\Theta^{*})\in M$ for which
$H(\xi^{*},\Theta^{*})=0$, then by definition 2, $\varepsilon_{0}$ will
be a bifurcation point.

Case 2. We assume that $H(\xi,\Theta)\neq 0$ at $\forall (\xi,\Theta)\in M$
and, hence, $\varepsilon_{0}$ is not a bifurcation point. Then
vector fields $H(\xi,0)$ and $H(\xi,1)$ are homotopic on the sphere  $\parallel\xi\parallel=r$.
Consequently, their rotations \cite{6} are coincided
$$
J(H(\xi,0),\parallel\xi\parallel=r)=J(H(\xi,1),(\parallel\xi\parallel=r) \eqno(12)
$$
Since vector fields  $H(\xi,0)$, $H(\xi,1)$ and their linearizations
$$
L^{-}_{1}(\xi)\stackrel{def}{=}\sum_{k=1}^{n}a_{ik}(\varepsilon(-\delta))\xi_{k}\mid_{i=1}^{n},
$$
$$
L^{+}_{1}(\xi)\stackrel{def}{=}\sum_{k=1}^{n}a_{ik}(\varepsilon(+\delta))\xi_{k}
\mid_{i=1}^{n}
$$
are nondegenerated on the sphere $\parallel\xi\parallel=r$, then by smallness of $r>0$,
fields $(H(\xi,0)$, $H(\xi,1)$ are homotopic to the linear parts $L^{-}_{1}(\xi)$
and $L_{1}^{+}(\xi)$.

Therefore
$$
J(H(\xi,0),\parallel\xi\parallel =r)=J(L_{1}^{-}(\xi),\parallel\xi\parallel =r)
\eqno(13)
$$
$$
J(H(\xi,1),\parallel\xi\parallel =r)=J(L_{1}^{+}(\xi),\parallel\xi\parallel =r).
\eqno(14)
$$
Because of  nondegeneracy of linear fields $L_{1}^{\pm}(\xi)$, by the theorem about
Kronecker index, the following equalities hold
$$
J(L^{-}_{1}(\xi),\parallel\xi\parallel=r)= sign\alpha(-\delta),
$$
$$
J(L_{1}^{+}(\xi),\parallel\xi\parallel=r)=sign\alpha(+\delta).
$$
Since $\alpha(-\delta)<0$, $\alpha(+\delta)>0$, then the equality (12)
is impossible by (13), (14). Hence, we find a pair $(\xi^{*},
\Theta^{*})\in M$ for which $H(\xi^{*},\Theta^{*})=0$ and $\varepsilon_{0}$
be a bifurcation point.

\begin{remark} If the conditions of the theorem 1 are satisfied for
$\forall\varepsilon\in\Omega_{0}\subset\Omega$, then $\Omega_{0}$ be a bifurcation
set of (6). If moreover, $\Omega_{0}$ is connected set and its every
point is contained in a neighbourhood, which is homeomorphic to some domain of
$R^{n}$, then $\Omega_{0}$ is called $n$-dimensional manifold of bifurcation.
\end{remark}

For example, it is true, if $\Upsilon=R^{n+1}$, $n\ge 1$, $\Omega_{0}$ be a
bifurcation set of (6) containing point  $\varepsilon=0$ and
$\nabla_{\varepsilon}\det[a_{ik}(\varepsilon)]\mid_{\varepsilon=0}\neq 0$.
It follows from the theorem 1 at $\Upsilon=R^{1}$ the generalization \cite{17},
and also other known strengthenings of M.A.~Krasnoselskii theorem on a bifurcation
point of odd multiplicity \cite{6}. 
An important results in the theory of bifurcation points were obtained for
(6) with potential BEq to $\xi$, when
$$
L(\xi,\varepsilon)=grad_{\xi}U(\xi,\varepsilon). \eqno(15)
$$
This condition is valid, if a matrix  $[\frac{\partial L_{k}}{\partial\xi_{i}}]_{
i,k=1}^{n}$ is symmetric. By differentiation of superposition, one finds from
(11) that
$$
\frac{\partial L_{k}}{\partial\xi_{i}}=<R_{x}\biggl(\sum_{s=1}^{n}\xi_{s}
\varphi_{s}+U(\xi,\varepsilon),\varepsilon\biggl)\biggl (\varphi_{i}+\frac
{\partial U}{\partial\xi_{i}}\biggl),\psi_{k}>, \eqno(16)
$$
where according to (8), (10)
$$
\varphi_{i}+\frac{\partial U}{\partial\xi_{i}}=(I-\Gamma R_{x})^{-1}\varphi_{i}.
\eqno(17)
$$
The operator $I-\Gamma R_{x}$ is continuously invertible because $\parallel\Gamma R_{x}
\parallel<1$ for sufficiently small by norm $\xi$ and $\varepsilon$. Substituting
(17) into (16) we obtain equalities
$$
\frac{\partial L_{k}}{\partial\xi_{i}}=<R_{x}(I-\Gamma R_{x})^{-1}\varphi_{i},\psi_{k}>,
\;i,k=1,\ldots,n.
$$
It follows the following claim:

\begin{lemma}
{ In order BEq (7) to be potential it is sufficient that a
matrix
$$
\Xi=[<R_{x}(\Gamma R_{x})^{m}\varphi_{i},\psi_{k}>]_{i,k=1}^{n}
$$
to be symmetric at $\forall (x,\varepsilon)$ in a neighbourhood of point (0,0)}.
\end{lemma}

\begin{corollary}
Let all matrices
$$
[<R_{x}(\Gamma R_{x})^{m}\varphi_{i},\psi_{k}>]_{i,k=1}^{n},\;\;m=0,1,2,\ldots
$$
are symmetric in some neighbourhood of point (0,0). Then BEq (7) be potential.
\end{corollary}

\begin{corollary}
 Let $E_{1}=E_{2}=H$, $H$ be Hilbert space.
If operator $B$ is symmetric in $D$, and operator $R_{x}(x,\varepsilon)$ is
symmetric for $\forall (x,\varepsilon)$ in a neighbourhood of point (0,0) in $D$,
then BEq be potential.
\end{corollary}

In the paper \cite{16}  more delicate sufficient conditions of BEq potentiality have been proposed.

Suppose that BEq (7) is potential. Then it follows from the proof of lemma 1 that the
corresponding potential $U$ in (15) has the form
$$
U(\xi,\varepsilon)=\frac{1}{2}\sum_{i,k=1}^{n}a_{i,k}(\varepsilon)\xi_{i}
\xi_{k}+\omega(\xi,\varepsilon),
$$
where $\parallel\omega(\xi,\varepsilon)\parallel=0(\mid\xi\mid^{2})$ at
$\xi\rightarrow 0$.

\begin{theorem}
{ Let BEq (7) be potential. Assume condition ${\bf A)}$.
Moreover, let the symmetrical matrix $[a_{ik}(\varepsilon(t))]$ possesses at least
$\nu_{1}$ positive eigenvalues at $t>0$ and at least $\nu_{2}$ positive
eigenvalues at $t<0$, $\nu_{1}\neq \nu_{2}$. Then $\varepsilon_{0}$
will be a bifurcation point of (6)}.
\end{theorem}

Proof. We take the arbitrary small $\delta>0$ and we consider the function
$U(\xi,\varepsilon ((2\Theta-1)\delta))$, defined at $\Theta\in[0,1]$
in a neighbourhood of the critical point $\xi=0$. \\
Case 1. If there is $\Theta^{*}\in[0,1]$ such that $\xi=0$ is the nonisolated critical
point of the function $U(\xi,\varepsilon((2\Theta^{*}-1)\delta)$, then by definition 2,
$\varepsilon_{0}$ will be a bifurcation point.\\
Case 2. Assume that point $\xi=0$ will be the isolated critical
point of the function $U(\xi,\varepsilon((2\Theta-1)\delta))$ at $\forall\Theta\in
[0,1]$, where $\varepsilon(t)$
be continuous function from condition {\bf A)}. Then at $\forall\Theta\in[0,1]$,
the Conley index \cite{2} $K_{\Theta}$ of the critical point $\xi=0$ of this function is
defined.
Let us remind that
$$
\det\parallel\frac{\partial^{2}U(\xi,\varepsilon((2\Theta-1)\delta))}{
\partial\xi_{i}\partial\xi_{k}}\parallel_{\xi=0}=\alpha((2\Theta-1)\delta).
$$
Since $\alpha((2\Theta-1)\delta)\neq 0$ at $\Theta\neq\frac{1}{2}$, then the
critical point  $\xi=0$ at $\Theta\neq\frac{1}{2}$ is nonsingular.
Therefore, index $K_{\Theta}$ for any $\Theta\neq\frac{1}{2}$ by the
definition  (here readers may refer to p.6 \cite{2}), is necessary equal to number of positive eigenvalues of
the corresponding Hessian. Thus, $K_{\Theta}=\nu_{1}$, $K_{1}=\nu_{2}$,
where $\nu_{1}\neq \nu_{2}$ by the condition of theorem 2.
Hence, $K_{\Theta}\neq K_{1}$. Suppose that $\varepsilon_{0}$ is not a bifurcation
point. Then $\nabla_{\xi}U(\xi,\varepsilon
((2\Theta-1)\sigma)\neq 0$ at $0<\parallel\xi\parallel\leq r$, where $r>0$
is small enough, $\Theta\in[0,1]$. Because of homotopic invariancy of Conley
index (see theorem 4, p.52 in \cite{2}), $K_{\Theta}$ is constant at $\Theta\in[0,1]$
and $K_{0}=K_{1}$. Hence, in the second case we find a pair  $(\xi^{*},
\Theta^{*})$ for arbitrary small $r>0$, $\delta>0$, where $0<\parallel\xi^{*}\parallel
\leq r$, $\Theta^{*}\in [0,1]$, satisfying to the equation $\nabla_{\xi}U(\xi,
\varepsilon((2\Theta-1)\delta)=0$ and $\varepsilon_{0}$ is a bifurcation point.

\begin{remark} Other proof of the theorem 2 with application of the Roll theorem
is given in \cite{18} for the case $\Upsilon=R^{1}$, $\nu_{+}=n$, $\nu_{-}=0$.
\end{remark}

\begin{remark} 
 The theorems 1, 2 (see remark 1) allow to construct not only the bifurcation
points, but also the bifurcation sets, surfaces and curves of bifurcation.
\end{remark}

\begin{corollary} Let $\Upsilon=R^{1}$ and BEq be potential. Moreover, let
$[a_{ik}(\varepsilon)]_{i,k=1}^{n}$ be positively definite matrix at $
\varepsilon\in (0,r)$ and negatively defined at  $\varepsilon\in(-r,0)$.
Then $\varepsilon=0$ is a bifurcation point of (6).
\end{corollary}

Consider the connection of eigenvalues of matrix $[a_{ik}(\varepsilon)]$
with eigenvalues of operator $B-R_{x}(0,\varepsilon)$.

\begin{lemma} {\it Let $E_{1}=E_{2}=E$, $\varepsilon\in R^{1}$; $\nu=0$ be
isolated Fredholm point of operator-function $B-\nu I$. Then
$$
sign\triangle(\varepsilon)=(-1)^{k} sign\prod_{i}^{k}\nu_{i}(\varepsilon)=
sign\prod_{i}^{n}\mu_{i}(\varepsilon),
$$
where $k$ be a root number of operator $B$; $\{\mu\}_{1}^{n}$
are eigenvalues of matrix} $[a_{ik}(\varepsilon)]$, $\triangle(\varepsilon)=\det
[a_{ik}(\varepsilon)]$.
\end{lemma}

Proof. Since $\{\mu_{i}\}_{1}^{n}$ are eigenvalues of matrix
$[a_{ik}(\varepsilon)]$, then $\prod_{i}^{n}\mu_{i}(\varepsilon)=
\triangle(\varepsilon)$. Thus, it is sufficient to prove the equality
$\triangle(\varepsilon)=(-1)^{k}\prod_{i}^{k}\nu_{i}(\varepsilon)$. Since zero is
the isolated Fredholm point of operator-function $B-\nu I$, then operators
$B$ and $B^{*}$ have the corresponding complete Jordan systems \cite{19}
$$
\varphi_{i}^{(s)}=(\Gamma)^{s-1}\varphi_{i}^{(1)},\;\;\psi_{i}^{(s)}=
(\Gamma^{*})^{s-1}\psi_{i}^{(1)}, \;\;i=1,\ldots,n;\;\;s=1,\ldots,P_{i}.
\eqno(18)
$$
Here
$$
<\varphi_{i}^{(P_{i})},\psi_{j}>=\delta_{ij};\;\;<\varphi_{i},\psi_{j}^{(P_{j})}>=
\delta_{ij},\;\;i,j=1,\ldots,n;\;\sum_{i=1}^{n}P_{i}=k.
$$
Let us remind that
$$
\varphi_{i}^{(1)}\stackrel{\triangle}{=}\varphi_{i}=\Gamma\varphi_{i}^{(P_{i})},
\;\;\psi_{i}^{(1)}
\stackrel{\triangle}{=}\psi_{i}=\Gamma^{*}\psi_{i}^{(P_{i})},\;\;\Gamma=\biggl (B+\sum_{1}^{n}<\cdot,
\psi_{i}^{(P_{i})}>\varphi_{i}^{(P_{i})}\biggl )^{-1}, \eqno(19)
$$
where $k=l_{1}+\ldots+l_{n}$ we call a root number of operator $B-R_{x}(0,\varepsilon)$.
The small eigenvalues $\nu(\varepsilon)$ of operator $B-R_{x}(0,\varepsilon)$
satisfy to the following branching equation \cite{19}
$$
L(\nu,\varepsilon)\stackrel{\triangle}{=}\det\mid<R_{x}(0,\varepsilon)+\nu I)(
I-\Gamma R_{x}(0,\varepsilon)-\nu\Gamma)^{-1}\varphi_{i},\psi_{j}>\mid_{i,j=1}^
{n}=0. \eqno(20)
$$
Because of preliminary Weierstrass theorem \cite{19}, p.~66, by the equalities (18),
(19), equation (20) in a neighbourhood of zero will be transformed to the form
$$
L(\nu,\varepsilon)\equiv(\nu^{k}+H_{k-1}(\varepsilon)\nu^{k-1}+\ldots +H_{0}
(\varepsilon))\Omega(\varepsilon,\nu)=0,
$$
where $H_{k-1}(\varepsilon),
\ldots,H_{0}(\varepsilon)=\triangle(\varepsilon)$ are continuous functions of $\varepsilon$,
$\Omega(0,0)\neq 0$, $H_{0}(0)=0$. Consequently, operator $B-R_{x}(0,\varepsilon)$
has $k\geq n$ small eigenvalues $\nu_{i}(\varepsilon)$, $i=1,\ldots,n$,
which we may define from the equation
$$
\nu^{k}+H_{k-1}(\varepsilon)\nu^{k-1}+\ldots+\triangle(\varepsilon)=0.
$$
Then $\prod_{i}^{k}\nu_{i}(\varepsilon)=\triangle(\varepsilon)(-1)^{k}$.

Assume now $\varepsilon\in R^{1}$. Consider the calculation of asymptotics of eigenvalues
$\mu(\varepsilon)$ and $\nu(\varepsilon)$. Let us introduce the block representation of
matrix $[a_{ik}]_{i,k=1}^{n}$, satisfying the following condition:

{\bf B)} Let $[a_{ik}(\varepsilon)]_{i,k=1}^{n}=[A_{ik}(\varepsilon)]_{i,k=1}^{l}
\sim[\varepsilon^{r_{ik}}A^{0}_{ik}]_{i,k=1}^{l}$ at $\varepsilon\rightarrow 0$,
where $[A_{ik}]$ are blocks of dimensionality $[n_{i}\times n_{k}]$, $n_{1}+
\ldots+n_{l}=n$,
$min(r_{i1},\ldots,r_{il})=r_{ii}\stackrel{\triangle}{=}r_{i}$ � $r_{ik}>r_{i}$
at $k>i$ (or at  $k<i$), $i=1,\ldots,l$. Let $\prod_{1}^{l}\det[A_{ii}^{0}]
\neq 0$.
The condition {\bf B)} means that matrix $[a_{ik}(\varepsilon)]_{i,k=1}^{n}$
admits the block representation being "asymptotically trianglar" at
$\varepsilon\rightarrow 0$.

\begin{lemma} { Assume {\bf B)}. Then
$$
\det[a_{ik}(\varepsilon)]_{i,k=1}^{n}=\varepsilon^{n_{1}r_{1}+\ldots+n_{l}r_{l}}
\biggl (\prod_{1}^{l}\det\mid A_{ii}^{0}\mid+0(1)\biggl ),
$$
formulas
$$
{\bf \mu}_{i}=\varepsilon^{r_{i}}({\bf C}_{i}+0(1)), \;\;i=1,\ldots,l \eqno(21)
$$
define the principal terms of all $n$ eigenvalues of matrix $\mid a_{ik}(\varepsilon)\mid_{i,k=1}^{n}$,
where ${\bf \mu}_{i}$, ${\bf C}_{i}\in R^{n_{i}}$; ${\bf C}_{i}$
be vector of eigenvalues of matrix $A_{ii}^{0}$}.
\end{lemma}

Proof. By {\bf B)} and the property of linearity of determinant, we have
$$
\det[a_{ik}(\varepsilon)]=\varepsilon^{n_{1}r_{1}+\ldots+n_{l}r_{l}}\det
\left|\begin{array}{lll}
A_{11}^{0}+0(1), & 0(1)\ldots\ldots, & \ldots\ldots 0(1) \\ [0.3cm]
A_{21}^{0}+0(1), & A_{22}^{0}+0(1), & 0(1)\ldots0(1) \\ [0.3cm]
\ldots\ldots\ldots & \ldots\ldots\ldots & \ldots\ldots\ldots  \\ [0.3cm]
A_{l1}^{0}+0(1), &\ldots\ldots, & A_{ll}^{0}+0(1)
\end{array}\right|=
$$
$$
\varepsilon^{n_{1}r_{1}+\ldots+n_{l}r_{l}}\biggl (\prod_{i}^{l}\det\mid A_{ii}^{0}
\mid+0(1)\biggl ).
$$
Substituting $\mu=\varepsilon^{r_{i}}c(\varepsilon)$, $i=1,\ldots,l$ into equation
$\det\mid a_{ik}(\varepsilon)-\mu\delta_{ik}\mid_{i,k=1}^{n}=0$
and using the property of linearity of determinant we obtain equation
$$
\varepsilon^{n_{1}r_{1}+\ldots+n_{i-1}r_{i-1}+(n_{i}+\ldots+n_{l})r_{i}}\{
\prod_{j=1}^{i-1}\det\mid A_{jj}^{0}\mid\cdot
$$
$$
\det(A_{ii}^{0}-�(\varepsilon)E)�(\varepsilon)^{n_{i+1}+\ldots+n_{l}}+
a_{i}(\varepsilon)\}=0,\;\;i=1,\ldots,l,  \eqno(22)
$$
where $a_{i}(\varepsilon)\rightarrow 0$ at $\varepsilon\rightarrow 0$.
Hence, the coordinates of unknown principal terms  ${\bf C}_{i}$ in asymptoticses
(21) satisfy
to the equations $\det\mid A_{ii}^{0}-�E\mid=0$, $i=1,\ldots,l$.

If $k=n$, then operator $B-R_{x}(0,\varepsilon)$, as well as the matrix
$[a_{ik}(\varepsilon)]_{i,k=1}^{n}$ has $n$ small eigenvalues. In this case we
state a result:

\begin{corollary} Let operator $B$ has not $I$- joined elements and let the condition
{\bf B)} holds. Then the formula
$$
{\bf \nu}_{i}=-\varepsilon^{r_{i}}({\bf C}_{i}+0(1)), \;\;i=1,\ldots,l,  \eqno(23)
$$
defines all $n$ small eigenvalues of operator $B-R_{x}(0,\varepsilon)$,
where ${\bf C}_{i}\in R^{n_{i}}$ be vector of eigenvalues of matrix $A_{ii}^{0}$,
$i=1,\ldots,l$, $n_{1}+\ldots+n_{l}=n$.
\end{corollary}

Proof. By lemma 2 in this case  $\sum_{1}^{n}P_{i}=n$
(root number $k=n$) and operator $B-R_{x}(0,\varepsilon)$ possesses at least $n$
small eigenvalues. Since $\sum_{1}^{l}n_{i}=n$, $A_{ii}^{0}$ is quadratic matrix,
then formula (23)
yields $n$ eigenvalues, where the principal terms coincide to within a sign with
principal terms in (21). For calculation of eigenvalues $\nu$ of operator
$B-R_{x}(0,\varepsilon)$ we transform (20) to the form
$$
L(\nu,\varepsilon)\equiv\det[a_{ik}(\varepsilon)+\sum_{j=1}^{\infty}b_{ik}^{(j)}
\nu^{j}]_{i,k=1}^{n}=0, \eqno(24)
$$
where
$$
b_{ik}^{(j)}=<[(I-\Gamma R_{x}(0,\varepsilon))^{-1}\Gamma]^{j-1}(I-\Gamma R_{x}(0,\varepsilon))
^{-1}\varphi_{i},\gamma_{k}>.
$$
Substituting $\nu=-\varepsilon^{r_{i}}c(\varepsilon)$ into (24) and taking into account
the property of linearity of determinant we shall receive the equation, which
differs from (22) by error term $a_{i}(\varepsilon)$ only.
Then in conditions of corollary 4 the principal terms of all small eigenvalues of operator
$B-R_{x}(0,\varepsilon)$ and matrix $-[a_{ik}(\varepsilon)]$
are defined from the same equations and therefore, are equal.

{\bf Conclusions.} 1) By lemma 3 we can replace condition {\bf A)}
in the theorem 1 with the following one:

${\bf A}^{*})$. Let $E_{1}=E_{2}=E$; $\nu=0$ be isolated Fredholm point of
operator-function $B-\nu I$. Let in a neighbourhood of point $\varepsilon_{0}
\in\Omega$ there is a set $S$, containing point $\varepsilon_{0}$ and be continuum
represented as $S=S_{+}\bigcup S_{-}$. Moreover, assume
$$
\varepsilon_{0}\in\partial S_{+}\bigcap\partial S_{-};\;\;\prod_{i}\nu_{i}
(\varepsilon)\mid_{\varepsilon\in S_{+}}\cdot\prod_{i}\nu_{i}(\varepsilon)
\mid_{\varepsilon\in S_{-}}<0,
$$
where $\{\nu_{i}(\varepsilon)\}$ are small eigenvalues of operator
$B-R_{x}(0,\varepsilon)$.

2) If the principal terms of asymptotics of small eigenvalues of operator
$B-R_{x}(0,\varepsilon)$ and matrix $[a_{ik}(\varepsilon)]_{i,k=1}^{n}$
coincide, then we may use eigenvalues of such operator in the theorem 2.
By corollary 4 it is possible, if $E_{1}=E_{2}=H$, operators
$B$ and $R_{x}(0,\varepsilon)$ are symmetric and condition $B)$ is valid.
Let us note that condition $B)$ is valid in papers \cite{15, 16, 17, 18} about bifurcation point
with potential BEq, thus $r_{1}=\ldots=r_{n}=1$.

\section{Statement of boundary-value problem and problem on a bifurcation
point for the system (32) \cite{9}}

We begin with a one preliminary result on reduction of VM system  (1)-(2)
with conditions (3) to the quasilinear system of elliptical equations for
distribution
(5), was first investigated in \cite{10}. Assume the following condition:

{\bf C).} $\hat{f}_{i}({\bf R},{\bf G})$ are fixed, differentiable functions in
distribution (5); $\alpha_{i}$, $d_{i}$ are free parameters;
$\mid d_{i}\mid\neq 0$; $\varphi_{i}=c_{1i}+l_{i}\varphi(r)$, $\psi_{i}=
c_{2i}+k_{i}\psi(r)$; $c_{1i}$, $c_{2i}$- const; the parameters $l_{i}$, $k_{i}$
are connected by relations
$$
l_{i}=\frac{m_{1}}{\alpha_{1}q_{1}}\frac{\alpha_{i}q_{i}}{m_{i}},\;\;
k_{i}\frac{q_{1}}{m_{1}}d_{1}=\frac{q_{i}}{m_{i}}d_{i},\;\;k_{1}=l_{1}=1,
\eqno(25)
$$
and the integrals $\int_{R^{3}}\hat{f}_{i}{\rm d}v$, $\int_{R^{3}}\hat{f}_{i}v{\rm d}v$
converge at $\forall\varphi_{i}$, $\psi_{i}$.

Let us introduce notations $m_{1}\stackrel{\triangle}{=}m$, $\alpha_{1}
\stackrel{\triangle}{=}\alpha$, $q_{1}\stackrel{\triangle}{=}q$.

\begin{theorem}
{ Let $f_{i}$ are defined as well as in (5) and the condition
${\bf C)}$ is valid. Let the vector-function $(\varphi,\psi)$ is a solution of the
system of equations
$$
\triangle\varphi=\mu\sum_{k=1}^{N}q_{k}\int_{R^{3}}f_{k}{\rm d}v, \;\;
\mu=\frac{8\pi\alpha q}{m}
$$
$$
\eqno(26)
$$
$$
\triangle\psi=\nu\sum_{k=1}^{N}q_{k}\int_{R^{3}}(v,d)f_{k}{\rm d}v, \;\;
\nu=-\frac{4\pi q}{mc^{2}}
$$
$$
\varphi\mid_{\partial D}=-\frac{2\alpha q}{m}u_{01}, \;\;\psi\mid_{\partial D}=
\frac{q}{mc}u_{02} \eqno(27)
$$
on a subspace
$$
(\partial_{r}\varphi_{i},d_{i})=0,\;\;(\partial_{r}\psi_{i},d_{i})=0,\;\;
i=1,\ldots,N. \eqno(28)
$$

Then the VM system (1), (2) with conditions (5) possesses a solution
$$
E=\frac{m}{2\alpha q}\partial_{r}\varphi, \;\;B=\frac{d}{d^{2}}(\beta+
\int_{0}^{1}(d\times J(tr),r){\rm d}t)-[d\times\partial_{r}\psi]
\frac{mc}{qd^{2}} \eqno(29),
$$
where
$$
J\stackrel{\triangle}{=}\frac{4\pi}{c}\sum_{k=1}^{N}q_{k}\int_{R^{3}}vf_{k}{\rm d}v,
\;\;\beta-const.
$$
The potentials
$$
U=-\frac{m}{2\alpha q}\varphi, \;\;A=\frac{mc}{qd^{2}}\psi d+A_{1}(r), \;\;
(A_{1},d)=0  \eqno(30)
$$
satisfying to condition (3) are defined through this solution}.
\end{theorem}

The proof of theorem 3 follows from theorem 1 of paper \cite{14}.

Introduce notations
$$
j_{i}=\int_{R^{3}}vf_{i}{\rm d}v, \;\;\rho_{i}=\int_{R^{3}}f_{i}{\rm d}v,
\;\;i=1,\ldots,N
$$
and the following condition:

{\bf D).} There are vectors  $\beta_{i}\in R^{3}$ such that $j_{i}=\beta_{i}
\rho_{i}$, $i=1,\ldots,N$.

For example, the condition  {\bf D} holds for distribution
$$
f_{i}=f_{i}(a(-\alpha_{i}v^{2}+\varphi_{i})+b((d_{i},v)+\psi_{i})) \eqno(31)
$$
for $\beta_{i}=\frac{b}{2\alpha_{i}a}d_{i}$, $a,b$-const.

Suppose that condition ${\bf D}$ is valid. Then the system (26) will be transformed to the
following
$$
\triangle\varphi=\lambda\mu\sum_{i=1}^{N}q_{i}A_{i},
\;\;\triangle\psi=\lambda\nu\sum_{i=1}^{N}q_{i}(\beta_{i},d)A_{i}, \eqno(32)
$$
where
$$
A_{i}(l_{i}\varphi,k_{i}\psi,\alpha_{i},d_{i})\stackrel{\triangle}{=}
\int_{R^{3}}\hat{f}_{i}{\rm d}v.
$$

Further, we shall suppose that the auxiliary vector $d$ in (5) is directed along
axes $Z$. Because of conditions (28) we put in system (32) $\varphi=\varphi(x,y)$,
$\psi=\psi(x,y)$, $x,y\in D\subset R^{2}$. Moreover, let $N\geq 3$ and
$\frac{k_{i}}{l_{i}}\neq const$.

Let $D$ be bounded domain in $R^{2}$ with the boundary $\partial D$ of class
$C^{2,\alpha}$, $\alpha\in (0,1]$. The boundary conditions (4) on the densities of  local
charge and current induce the equalities:

{\bf I.}
$$
\sum_{k=1}^{N}q_{k}A_{k}(l_{k}\varphi^{0}, k_{k}\psi^{0},\alpha_{i},d_{i})=0;\;\;
\sum_{k=1}^{N}q_{k}(\beta_{k},d)A_{k}(l_{k}\varphi^{0},k_{k}\psi^{0},\alpha_{i},
d_{i})=0  \eqno(33)
$$
for $\forall\varepsilon\in \iota$, where $\iota$ is a neighbourhood of point
$\varepsilon=0$
and
$$
\varphi^{0}=-\frac{2\alpha q}{m}u_{01},\;\;\psi^{0}=\frac{q}{mc}u_{02}.
\eqno(34)
$$

\begin{remark} If $N=2$ and $\beta_{i}=\frac{d_{i}}{2\alpha_{i}}$, then by
condition {\bf I} and equalities $(\beta_{i},d)=\frac{d^{2}}{2\alpha}\frac{k_{i}}
{l_{i}}$ we have alternative: or in condition {\bf I}: $A_{1}=A_{2}=0$ or $k_{i}=
l_{i}$, $i=1,2$. In this case, and also at $\frac{k_{i}}{l_{i}}$=const the system
(32) is reduced to one equation and bifurcation of solutions in such approach, as it
is considered in this paper is impossible.
\end{remark}

By (33), (34) system (32) with boundary conditions
$$
\varphi\mid_{\partial D}=\varphi^{0}, \;\;\psi\mid_{\partial D}=\psi^{0} \eqno(35)
$$
has a trivial solution  $\varphi=\varphi^{0}$, $\psi=\psi^{0}$ at
$\forall\lambda\in R^{+}$.

Then because of theorem 3 the VM system with boundary conditions (3), (4)
has a trivial solution at $\forall\lambda$
$$
E^{0}=\frac{m}{2\alpha q}\partial_{r}\varphi^{0}=0, \;\;B^{0}=\beta d_{1}, \;\;
r\in D\subset R^{2},
$$
$$
f^{0}=\lambda\hat{f}_{i}(-\alpha_{i}v^{2}+c_{1i}+l_{i}\varphi^{0},\;(v,d_{i})+
c_{2i}+k_{i}\psi^{0}).
$$
Thus, the densities $\rho$ and $j$ vanish at domain $D$.

Now our purpose is to find $\lambda_{0}$ in neighbourhood of which system (32), (35)
has a nontrivial solution. Then the corresponding densities $\rho$ and $j$
will be identically vanish at domain  $D$, and the point  $\lambda_{0}$ is a
bifurcation point of the VM system with conditions (4), (5).

Let functions $f_{i}$ are analytical in (5). Using the expansion in Taylor series
$$
A(x,y)=\sum_{i\geq 0}^{\infty}\frac{1}{i!}((x-x^{0})\frac{\partial}{\partial x}+
(y-y^{0})\frac{\partial}{\partial y})^{i}A(x^{0},y^{0})
$$
and selecting linear terms, we transform (32) to operator form
$$
(L_{0}-\lambda L_{1})u-\lambda  r(u)=0. \eqno(36)
$$
Here
$$
L_{0}=\left[ \begin{array}{cc}
\triangle & 0 \\ [0.3cm]
0 & \triangle
\end{array}
\right],\;\;u=(\varphi-\varphi^{0},\;\psi-\psi^{0})'; \eqno(37)
$$
$$
L_{1}=\sum_{s=1}^{N}q_{s}\left[ \begin{array} {cc}
\mu l_{s}\frac{\partial A_{s}}{\partial x} & \mu k_{s}\frac{\partial A_{s}}
{\partial y}\\ [0.5cm]
\nu l_{s}(\beta_{s},d)\frac{\partial A_{s}}{\partial x} & \nu k_{s}(\beta_{s},d)
\frac{\partial A_{s}}{\partial y}
\end{array}
\right ]_{x=l_{s}\varphi^{0},y=k_{s}\psi^{0}}\stackrel{\triangle}{=}
$$
$$
\left [\begin{array}{cc}
\mu T_{1} & \mu T_{2}\\ [0.5cm]
\nu T_{3} & \nu T_{4}
\end{array}
\right ]; \eqno(38)
$$
$$
r(u)=\sum_{i\geq l}^{\infty}\sum_{s=1}^{n}\varrho_{is}(u)b_{s}, \eqno(39)
$$
where
$$
\varrho_{is}(u)\stackrel{\triangle}{=}\frac{q_{s}}{i!}(L_{s}u_{1}\frac{\partial}
{\partial x}+k_{s}u_{2}\frac{\partial}{\partial y})^{i}A_{s}(l_{s}\varphi^{0},
k_{s}\psi^{0})
$$
are $i$ homogeneous forms by $u$;
$$
\frac{\partial^{i_{1}+i_{2}}}{\partial x^{i_{1}}\partial y^{i_{2}}}A_{s}(x,y)
\mid_{x=l_{s}\varphi^{0},y=k_{s}\psi^{0}}=0 \;\;{\rm at}
$$
$$
2\leq i_{1}+i_{2}\leq l-1, \;\;s=1,\ldots,N;\;\;l\geq 2;\;\;
b_{s}\stackrel{\triangle}{=}(\mu,\nu(\beta_{s},d))'.
$$
We study the problem of existence of a bifurcation point $\lambda^{0}$
for (32), (34) as the problem on bifurcation point for operator equation (36).
Let us introduce Banach spaces $C^{2,\alpha}(\bar{D})$ and $C^{0,\alpha}(\bar{D})$
with norms $\parallel\cdot\parallel_{2,\alpha}$, $\parallel\cdot
\parallel_{0,\alpha}$ and $W^{2,2}(D)$, which is usual $L^{2}$ Sobolev space
in $D$.
Let us introduce  Banach space $E$ of vectors $u\stackrel{\triangle}{=}(u_{1},u_{2})'$,
where $u_{i}\in L_{2}(D)$, $L_{2}$ be real Hilbert space with internal product
$(\;,\;)$ and the cor\-respon\-ding norm $\parallel\cdot
\parallel_{L_{2}}(D)$. As a range of definition $D(L_{0})$ we take set of vectors
$u\stackrel{\triangle}{=}(u_{1},u_{2})$ with $u_{i}\in\stackrel{\circ}{
W}^{2,2}(D)$. Here $\stackrel{\circ}{W}^{2,2}(D)$ denotes $W^{2,2}$ functions
with trace 0 on $\partial D$. Hence, $L_{0}: D\subset E\rightarrow E$ is linear
self-adjoint operator. By virtue of embedding
$$
W^{2,2}(D)\subset C^{0,\alpha}(\bar{D}), \;\;0<\alpha<1 \eqno(40)
$$
the operator $r: W^{2,2}\subset E\rightarrow E$ be analytical in neighbourhood of zero.
The operator $L_{1}\in L(E\rightarrow E)$ is linear bounded. For matrix
corresponding to operator $L_{1}$ we shall keep same notations. By embedding (40)
any solution of the equation (36) will be H$\ddot{o}$lder in $D(L_{0})$.
Moreover, because the coefficients of (36) are constant, then vector $r(u)$
will be analytical,
$\partial D\in C^{2,\alpha}$ and thanks to well-known results of the regularity theory
of weak solutions [8], the being searching generalized solutions
of (36) in $\stackrel{\circ}{W}^{2,2}(D)$ belong to $C^{2,\alpha}$.
By theorem 3 on reduction of VM system, the bifurcation points of problem (32), (34)
are the bifurcation points of solutions of VM system (1), (2)
with boundary conditions (3), (4).
Thanks to given conditions on $L_{0}$ and $L_{1}$, all singular points of operator-
function $L(\lambda)\stackrel{\triangle}{=} L_{0}-\lambda L_{1}$ be Fredholm.
The bifurcation points of nonlinear equation (36) we can found only among points of
a spectrum for linearized system
$$(L_{0}-\lambda L_{1})u=0. \eqno(41)$$
For study of spectrum problem (41) we preliminary find the eigenvalues and the eigenfunctions
of matrix $L_{1}$ in (41) for physically admissible parameters.
With this purpose, we introduce the following condition:

{\bf II:} $(T_{1}T_{4}-T_{2}T_{3})>0$, $\;T_{1}<0$.

\begin{lemma}
{Let $\frac{\partial A_{i}}{\partial x}=\frac{\partial A_{i}}
{\partial y}>0$, $i=1,\ldots,N$ at $x=l_{i}\varphi^{0}$, $y=k_{i}\psi^{0}$.
Assume
$$
\sum_{i=2}^{N}\sum_{j=1}^{i-1}a_{i}a_{j}(l_{j}k_{i}-k_{j}l_{i})(\beta_{i}-
\beta_{j}, d)>0,
$$
where $a_{i}\stackrel{\triangle}{=}q_{i}\frac{\partial A_{i}}{\partial x}$, then
condition  {\bf II} is valid}.
\end{lemma}

Proof. Without loss of generality we put $q\stackrel{\triangle}{=}q_{1}<0$,
$q_{i}>0$, $i=2,\ldots,N$. Then via (25) $sign q_{i}l_{i}=sign q$.
Further, because of definition of $T_{1}$ (see.(38)), we verify that $T_{1}<0$.
The positiveness of $T_{1}T_{4}-T_{2}T_{3}$ follows from equality
$$
T_{1}T_{4}-T_{2}T_{3}=\sum l_{i}a_{i}\sum k_{i}(\beta_{i},d)a_{i}-
\sum k_{i}a_{i}\sum l_{i}(\beta_{i},d)a_{i}=
$$
$$
\sum_{i=2}^{N}\sum_{j=1}^{i-1}a_{i}a_{j}(l_{j}k_{i}-k_{j}l_{i})(\beta_{i}-
\beta_{j},d).
$$

\noindent{\bf Example.} 
If $\beta_{i}=\frac{d_{i}}{2\alpha_{i}}$, then $(\beta_{i},d)=
\frac{d^{2}}{2\alpha}\frac{k_{i}}{l_{i}}$ and
$$
\sum_{i=2}^{N}\sum_{j=1}^{i-1}=a_{i}a_{j}(l_{j}k_{i}-l_{i}k_{j})^{2}\cdot
\frac{d^{2}}{2\alpha l_{i}l_{j}}>0.
$$

\begin{lemma}
{\it Let distribution function has a form (31) and
$f^{'}_{i}>0$. Then conditions {\bf D} and {\bf II}
hold for $\beta_{i}=\frac{b}{a}\frac{d_{i}}{2\alpha_{i}}$, and the system (32)
will be transformed to the potential form
$$
\triangle \left [\begin{array}{c}
\varphi \\ [0.5cm]
\psi
\end{array}
\right ]=\lambda \left [ \begin{array}{cc}
a_{1} & 0 \\ [0.5cm]
0 & a_{2}
\end{array}
\right ]
\left [ \begin{array}{c}
\frac{\partial V}{\partial \varphi} \\ [0.5cm]
\frac{\partial V}{\partial \psi}
\end{array}
\right ], \eqno(42)
$$
where}
$$
V=\sum_{k=1}^{N}\frac{q_{k}}{l_{k}}\int_{0}^{al_{k}\varphi+
bk_{k}\psi}A_{k}(s)ds, \;\;a_{1}=\mu/a, \;\;a_{2}=\frac{\nu d^{2}}{2ab}. \eqno(43)
$$
\end{lemma}
The proof is conducted by direct substitution (43) into the system (42).

\begin{lemma}
{\it Let $r\stackrel{\triangle}{=}x\in R^{1}$, $v\in R^{2}$,
$d\stackrel{\triangle}{=}d_{2}$. Then the system (32) with potential (43)
can be written as Hamiltonian system
$$
\dot{p}_{\varphi}=-\partial_{\varphi}H, \;\;\;\dot{\varphi}=\partial_{p_{\varphi}}
H
$$
$$
\dot{p}_{\psi}=-\partial_{\psi}H, \;\;\;\dot{\psi}=\partial_{p_{\psi}}H
$$
with Hamiltonian function of the form
$$
H=-\frac{p_{\varphi}^{2}}{2}-\frac{p_{\psi}^{2}}{2}+V(\varphi(x),\psi(x)).
$$
Here}
$$
V(\varphi,\psi)=\lambda a_{1}\sum_{k=1}^{N}\frac{q_{k}}{l_{k}}\int_{0}^{al_{k}
\varphi}\int_{R^{2}}A(s,\psi)ds+\lambda a_{2}\sum_{k=1}^{N}\frac{q_{k}}{l_{k}}
\int_{0}^{bk_{k}\psi}\int_{R^{2}}A(\varphi,s)ds.
$$
\end{lemma}
The proof follows from lemma 2.2 (p.1152) of work \cite{4}.

\begin{lemma}
{ Assume {\bf II}. Then matrix $L_{1}$ in
(38) has one positive eigenvalue
$$
\chi_{+}=\mu T_{1}+0(1)
$$
and one negative
$$
\chi_{-}=\eta\frac{T_{1}T_{4}-T_{2}T_{3}}{T_{1}}\epsilon+O(\epsilon),\;\;\;
\eta=\frac{4\pi\mid q\mid}{m}>0  \eqno(44)
$$
at $\epsilon\stackrel{\triangle}{=}\frac{1}{c^{2}}\rightarrow 0$.

Eigenvalue $\chi_{-}$ induces the eigenvactors of matrices
$L_{1}$ and $L_{1}'$ respectively}
$$
\left[\begin{array}{c}
c_{1} \\ [0.5cm]
c_{2}
\end{array}
\right]=\left[ \begin{array}{c}
-\frac{T_{2}}{T_{1}}\\ [0.5cm]
0
\end{array}
\right]+O(\epsilon), \;\;\;
\left[ \begin{array}{c}
c^{*}_{1}\\ [0.5cm]
c^{*}_{2}
\end{array}
\right]=
\left[ \begin{array}{c}
0 \\ [0.5cm]
1
\end{array}
\right]+O(\epsilon).
$$
\end{lemma}

The readers may refer to \cite{14} for the proof.

Let us now consider the calculation of bifurcation points $\lambda_{0}$ of equation (36).
Setting in (36) $\lambda=\lambda_{0}+\epsilon$, we consider the equation
$$
(L_{0}-(\lambda_{0}+\epsilon)L_{1}u-(\lambda_{0}+\epsilon)r(u)=0 \eqno(45)
$$
in neighbourhood of point $\lambda_{0}$. Let $T_{2}\neq 0$ and $T_{3}\neq 0$, or
$T_{2}=T_{3}=0$. With the purpose of symmetrization of system at $T_{2}\neq 0$
and $T_{3}\neq 0$ having multiplicated both parts of (45) on matrix
$$
M=\left(\begin{array}{cc}
1 & 0 \\ [0.5cm]
0 & \tilde{a}
\end{array}
\right), \;\;{\rm where}\;\;\tilde{a}\stackrel{\triangle}{\equiv}\frac{\mu T_{2}}{
\nu T_{3}}\neq 0,
$$
we write (45) as
$$
Bu=\epsilon B_{1}u+(\lambda_{0}+\epsilon)\Re(u).  \eqno(46)
$$
Here $B=M(L_{0}-\lambda_{0}L_{1})$; $\Re(u)\stackrel{\triangle}{=}Mr(u)\stackrel
{\triangle}{=}(r_{1}(u),r_{2}(u))$; $B_{1}\in L(E\rightarrow E)$ be Fredholm self-
adjoint operator. If $A_{s}=A_{s}(al_{s}\varphi+bk_{k}\psi)$, then
$$
\frac{\partial A_{s}}{\partial y}=A'_{s}b,\;\;\frac{\partial A_{s}}{\partial x}=
A'_{s}a, \;\;\tilde{a}=\mu b/(\nu\frac{d^{2}}{2\alpha a}), \;\;\beta_{s}=
\frac{b}{a}\frac{d_{s}}{2\alpha_{s}}.
$$
In expansion (39)
$$
\varrho_{is}=\frac{q_{s}}{i!}A^{(i)}_{s}(al_{s}\varphi^{0}+bk_{s}\psi^{0})(
al_{s}u_{1}+bk_{s}u_{2})^{i}.
$$
Thus, in this case $\frac{\partial r_{1}}{\partial u_{2}}=\frac{\partial r_{2}}
{u_{1}}$ matrix $\Re_{u}(u)$ will be symmetric for $\forall u$ and operator
$\Re_{u}: E\rightarrow E$ is self-adjoint for $\forall u$.

\begin{remark} If $T_{2}=T_{3}=0$, then we put $\tilde{a}=1$. If
$T_{2}=0$, $T_{3}\neq 0$ or $T_{3}=0$, $T_{2}\neq 0$, then the problem (36)
has not the property of symmetrization and we should work with (45). In
this case for study of the problem on bifurcation point we may use our results from
[13].
\end{remark}

Let $\mu$ be eigenvalue of the Dirichlet problem
$$
-\triangle e=\mu e \;\;e\mid_{\partial D}=0 \eqno(47)
$$
and $\{e_{1},\ldots,e_{n}\}$ be orthonormalized basis in a subspace of eigenfunctions.
Denote by $c_{-}=(c_{1},c_{2})'$ the eigenvector of matrix
$L_{1}$, which corresponds to eigenvalue $\chi_{-}<0$.

{\bf Lemma 8.} {\it Let $\lambda_{0}=-\mu/\chi_{-}$. Then $\lambda_{0}>0$,
$dim N(B)=n$ and the system $\{{\bf e}_{i}\}_{i=1}^{n}$, where ${\bf e}_{i}=c_{-}e_{i}$
forms basis in a subspace} $N(B)$.

Proof. Let us introduce matrix of columns $\Lambda$, which are the eigenvectors of
matrix $L_{1}$ corresponding to eigenvalues $\chi_{-}$,
$\chi_{+}$. Moreover,
$$
\Lambda^{-1}L_{1}\Lambda=\left(\begin{array}{cc}
\chi_{-} & 0 \\ [0.5cm]
0 & \chi_{+}
\end{array}
\right), \;\;L_{0}\Lambda=\Lambda L_{0}
$$
and equation $Bu=0$ by change $u=\Lambda U$ will be transformed to the form
$$
M[L_{0}\Lambda U-\lambda_{0}L_{1}\Lambda U]=M[\Lambda(L_{0}U-\lambda_{0}\Lambda^{-1}
L_{1}\Lambda U)]=0.
$$
Hence, from here follows that the linear system (41) is decomposed onto two
linear elliptical equations
$$
\triangle U_{1}-\lambda_{0}\chi_{-}U_{1}=0, \;\;U_{1}\mid_{\partial D}=0, \;\;\;
\triangle U_{2}-\lambda_{0}\chi_{+}U_{2}=0,\;\;U_{2}\mid_{\partial D}=0, \eqno(48)
$$
where  $\lambda_{0}\chi_{-}=-\mu$, $\lambda_{0}\chi_{+}>0$. From  (47) follows that
$\mu\in\sigma(-\triangle)$. Hence, $U_{1}=\sum_{i=1}^{n}\alpha_{i}e_{i}$,
$\alpha_{i}-const$, $U_{2}=0$ and
$$
\left|\begin{array}{c}
u_{1} \\ [0.5cm]
u_{2}
\end{array}
\right|=\Lambda U=\left|\begin{array}{cc}
c_{1-} & c_{1+} \\ [0.5cm]
c_{2-} & c_{2+}
\end{array}
\right|=\left|\begin{array}{c}
U_{1} \\ [0.5cm]
0
\end{array}
\right|=\left|\begin{array}{c}
c_{1-} \\ [0.5cm]
c_{2-}
\end{array}
\right|\sum_{i=1}^{n}\alpha_{i}e_{i}.
$$
Let us construct Lyapunov-Schmidt BEq for equation (46).

Without loss of generality we assume that the eigenvector $c_{1-}$ of matrix
$L_{1}$ is chosen such that $\chi_{-}(c_{1-}^{2}+Fc_{2-}^{2})=1$, where
$F=\frac{\mu T_{2}}{\nu T_{3}}$. Then the system of vectors $\{B_{1}{\bf e}_{i}\}_{i=1}^{n}$
is biorthogonal to $\{{\bf e}_{i}\}_{i=1}^{n}$. Thus, operator
$$
\breve{B}=B+\sum_{1}^{n}<\cdot,\gamma_{i}>\gamma_{i}
$$
with $\gamma_{i}\stackrel{\triangle}{=}B_{1}{\bf e}_{i}$ has inverse bounded
$\Gamma\in L(E\rightarrow E)$, $\Gamma=\Gamma^{*}$, $\Gamma\gamma_{i}=
{\bf e}_{i}$.

Rewrite (46) as the system
$$
(\breve{B}-\epsilon B_{1})u=(\lambda_{0}+\epsilon)\Re(u)+\sum_{i}\xi_{i}\gamma_{i}
\eqno(49)
$$
$$
\xi_{i}=<u,\gamma_{i}>,\;\;i=1,\ldots,n. \eqno(50)
$$
By the theorem on inverse operator we have from (49)
$$
u=(\lambda_{0}+\epsilon)(I-\epsilon\Gamma B_{1})^{-1}\Gamma\Re(u)+
\frac{1}{1-\epsilon}\sum_{i=1}^{n}\xi_{i}{\bf e}_{i}. \eqno(51)
$$
From (50) we have
$$
\frac{\epsilon}{1-\epsilon}\xi_{i}+\frac{\lambda_{0}+\epsilon}{1-\epsilon}
<\Re(u),{\bf e}_{i}>=0, \eqno(52)
$$
where $\Re(u)=\Re_{l}(u)+\Re_{l+1}(u)+\ldots.$ Because of the theorem on
implicit operator, equation (51) has unique solution for sufficiently small
 $\epsilon$, $\mid\xi\mid$.
$$
u=u_{1}(\xi{\bf e},\epsilon)+(\lambda_{0}+\epsilon)(I-\epsilon\Gamma B_{1})^{-1}\Gamma
\{u_{l}(\xi{\bf e},\epsilon)+u_{l+1}(\xi {\bf e},\epsilon)+\ldots\}. \eqno(53)
$$
Here
$$
u_{1}(\xi{\bf e},\epsilon)=\frac{1}{1-\epsilon}\sum_{i=1}^{n}\xi_{i}{\bf e}_{i},
$$
$$
u_{l}(\xi{\bf e},\epsilon)=\Re_{l}(u_{1}(\xi{\bf e},\epsilon)),
$$
$$
u_{l+1}(\xi{\bf e},\epsilon)=\Re_{l+1}(u_{1}(\xi{\bf e},\epsilon))+
$$
$$
+\left\{\begin{array}{ll}
0, & l\geq 2 \\ [0.5cm]
\Gamma\Re_{2}'(u_{1}(\xi{\bf e},\epsilon))(\lambda_{0}+\epsilon)(I-\epsilon
\Gamma B_{1})^{-1}\Gamma u_{2}(\xi{\bf e},\epsilon), & l=2
\end{array}
\right.
$$
and etc.
Substituting the solution  (53) into (52) we obtain desired BEq
$$
{\bf L}(\xi,\epsilon)=0 \eqno(BEq)
$$
with ${\bf L}=(L^{1},\ldots,L^{n})$,
$$
{\bf L}^{i}=\frac{\epsilon}{1-\epsilon}\xi_{i}+\frac{\lambda_{0}+\epsilon}{
(1-\epsilon)^{l+1}}[<\Re_{l}(\xi{\bf e},{\bf e}_{i})>+\frac{1}{1-\epsilon}<
\Re_{l+1}(\xi{\bf e},{\bf e}_{i})>]+
$$
$$
\left\{ \begin{array}{l}
0,\;\;\;\;\;  l>2 \\ [0.5cm]
\frac{\lambda_{0}+c}{(1-\varepsilon)^{4}}<\Re_{2}^{'}(\xi\hbox{\bf e}(I-\varepsilon
\Gamma B_{1})^{-1}\Gamma\Re_{2}(\xi\hbox{\bf e}),\hbox{\bf e}_{i}>, \;\;\;\;\; l=2
\end{array}
\right.+
r_{i}(\xi,\varepsilon),
$$
$r_{i}=o(\mid\xi\mid^{l+1})$, $i=1,\ldots,n$. If ${\bf L}(\xi,\varepsilon)=
grad \,U(\xi,\varepsilon)$, then we call BEq potential. In potential case matrix
${\bf L}_{\xi}(\xi,\varepsilon)$ is symmetric.

Let in (46) $f_{i}=f_{i}(al_{i}\varphi+bk_{i}\psi)$, $i=1,\ldots,N$. Then
from explained above matrix $\Re_{u}(u)$ will be symmetric at $\forall u$ and
we have the following statement:

\begin{lemma} 
{Let conditions {\bf C)}, {\bf D)}, {\bf I-II} and
$\lambda_{0}=-\mu/\chi_{-}$ hold. Then equation (46) possesses so much small solutions
$u\rightarrow 0$ at $\lambda\rightarrow\lambda_{0}$, as small solutions $\xi
\rightarrow 0$ possesses BEq at $\varepsilon\rightarrow 0$. If in system (32)
$A_{i}=A_{i}(al_{i}\varphi+bk_{i}\psi)$, $i=1,\ldots,N$; $a, b-$-const, then
BEq will be potential}.
\end{lemma}

\begin{theorem}[{\bf Principal theorem}] { Let $N\geq 3$. Let conditions
{\bf C}, {\bf D}, {\bf I-II} and $\lambda_{0}=-\mu/\chi$ are valid, where
 $\mu$ is $n$ multiple eigenvalue of Dirichlet problem (47). Number $\chi_{-}$
see in (44). If $n$ is odd, or distribution function has the form
$f_{i}=f_{i}(a(-\alpha_{i}v^{2}+\varphi_{i})+b((d_{i},v)+\psi_{i}))$, $i=1,\ldots,
N$, then $\lambda_{0}$ be a bifurcation point of VM system �� (1)-(2)
with conditions} (3)-(4).
\end{theorem}

Proof. Case 1. Let $n$ is odd. Then in BEq
$$
\triangle(\varepsilon)\equiv\det \left|\frac{\partial L_{k}}{ \partial \xi_{i}}(0,\varepsilon)
\right|^{n}_{i,k=1}=\left (\frac{\varepsilon}{1-\varepsilon}\right)^{n}.
$$
Since $n$ is odd, then $\triangle(\varepsilon)>0$ for $\varepsilon\in(0,1)$,
and $\triangle(\varepsilon)<0$ for $\varepsilon\in(-1,0)$ and the statement of theorem
follows from theorem 1.

Case 2. Let $f_{i}=f_{i}(a(-\alpha_{i}v^{2}+\varphi_{i})+b((d_{i},v)+\psi_{i}))$.
Then BEq is potential, moreover
$$
\frac{\partial L_{k}(0,\varepsilon)}{\partial \xi_{i}}=\frac{\varepsilon}{1-\varepsilon}
\delta_{ik}, \;\;i,k=1,\ldots,n.
$$
Hence, all eigenvalues of matrix $\parallel\frac{\partial L_{k}(0,\varepsilon)}
{\partial\xi_{i}}\parallel$ are positive at $\varepsilon>0$ and are negative at
$\varepsilon<0$. Thus, the validity of the theorem in case 2 follows from theorem 2.

\section{Conclusion}
The distributions functions $f_{i}$ in VM system
depend not only upon  $\lambda$, but also on parameters $\alpha_{i}$, $d_{i}$, $k_{i}$, $l_{i}$.
It seems interest to investigate a behaviour of solutions of (1)-(2) with
conditions (3), (4) depending from these parameters. Applying theorems 1, 2 and
their corollaries in the present paper, we can prove the existence theorems of points and
surfaces of bifurcation for this more complicated case.


\end{document}